\documentclass{Interspeech}

\interspeechcameraready

\title{DnR-nonverbal: Cinematic Audio Source Separation Dataset\\Containing Non-Verbal Sounds}

\author[affiliation={1}]{Takuya}{Hasumi}
\author[affiliation={1}]{Yusuke}{Fujita}

\affiliation{}{LY Corporation}{Japan}
\email{takuya.hasumi@lycorp.co.jp, yusuke.fujita@lycorp.co.jp}

\usepackage{comment}

\usepackage{cite}
\usepackage{amsmath,amssymb,amsfonts,bm}
\usepackage{algorithm}
\usepackage{algpseudocode}
\usepackage{graphicx}
\usepackage{textcomp}
\usepackage{caption}
\usepackage{subcaption}
\usepackage{multirow}
\usepackage{xcolor}

\def\ReadingSet{\mathcal{R}}
\def\NonverbalSet{\mathcal{N}}
\def\speech{\mathrm{s}}
\def\verbal{\mathrm{v}}
\def\reading{\mathrm{r}}
\def\nonverbal{\mathrm{n}}
\def\ZTP{\mathcal{Z}\mathcal{T}\mathcal{P}}
\def\SkewNorm{\tilde{\mathcal{G}}}

\begin{document}

\maketitle

\begin{abstract}
We propose a new dataset for cinematic audio source separation (CASS) that handles non-verbal sounds.
Existing CASS datasets only contain reading-style sounds as a speech stem.
These datasets differ from actual movie audio, which is more likely to include acted-out voices.
Consequently, models trained on conventional datasets tend to have issues where emotionally heightened voices, such as laughter and screams, are more easily separated as an effect, not speech.
To address this problem, we build a new dataset, DnR-nonverbal.
The proposed dataset includes non-verbal sounds like laughter and screams in the speech stem.
From the experiments, we reveal the issue of non-verbal sound extraction by the current CASS model and show that our dataset can effectively address the issue in the synthetic and actual movie audio.
Our dataset is available at \url{https://zenodo.org/records/15470640}.
\end{abstract}

\section{Introduction}
Cinematic audio source separation (CASS)~\cite{uhlich2024sound} aims to decompose the movie audio into sources.
Typically, this task defines speech, music, and effects as the exclusive target stems.
The CASS helps restore old movies and analyze movie content by demixing the audio.
The technique may also be applicable to detect copyrighted music from audio in advertisement videos.

Thanks to the recent development of deep learning techniques in speech separation~\cite{wang2018supervised,luo2019conv,wang2023tf}, music source separation~\cite{stoter20182018,stoter2019open,rouard2022hybrid}, and universal source separation~\cite{8937253}, CASS also utilizes the deep neural network-based model.
Recently, the pair of stem-shared encoder and stem-wise decoder, such as MRX~\cite{petermann2022cocktail} and BandIt~\cite{watcharasupat2023generalized}, has been commonly utilized in CASS.
In MRX, the model encodes multi-resolution amplitude spectrograms and decodes the encoded features to estimate multi-resolution amplitude masks.
The model utilizes acoustic features with high temporal resolution and features with fine frequency resolution.
BandIt, one of the state-of-the-art CASS models, encodes complex spectrograms and decodes them to estimate complex spectrogram masks.
The model uses an efficient temporal and frequency modeling network by leveraging band-split RNN~\cite{luo2023music}.

One practical issue of these CASS models is that they fail to separate expressive speeches, typically sounds containing non-verbal sounds, such as laughter and screaming.
Though humans utter these sounds, they are separated as the effect stem by the CASS models, as we will show in Sec.~\ref{sec:experiments/performance}.
The root cause of this issue is that the conventional CASS datasets contain only reading-style speeches and exclude expressive non-verbal sounds.
In the conventional CASS datasets, speech tracks are collected from ASR corpora.
Specifically, the widely known Divide and Remaster v2 (DnR-v2)~\cite{petermann2022cocktail} uses LibriSpeech~\cite{panayotov2015librispeech}.
The DnR-v3 extends DnR-v2 by collecting multiple ASR corpora to support multi-lingual speeches.
The domain mismatch between the synthetic dataset and realistic cinematic audio causes the undesired behavior of the CASS model.

\begin{figure}
    \centering
    \includegraphics[width=0.85\linewidth]{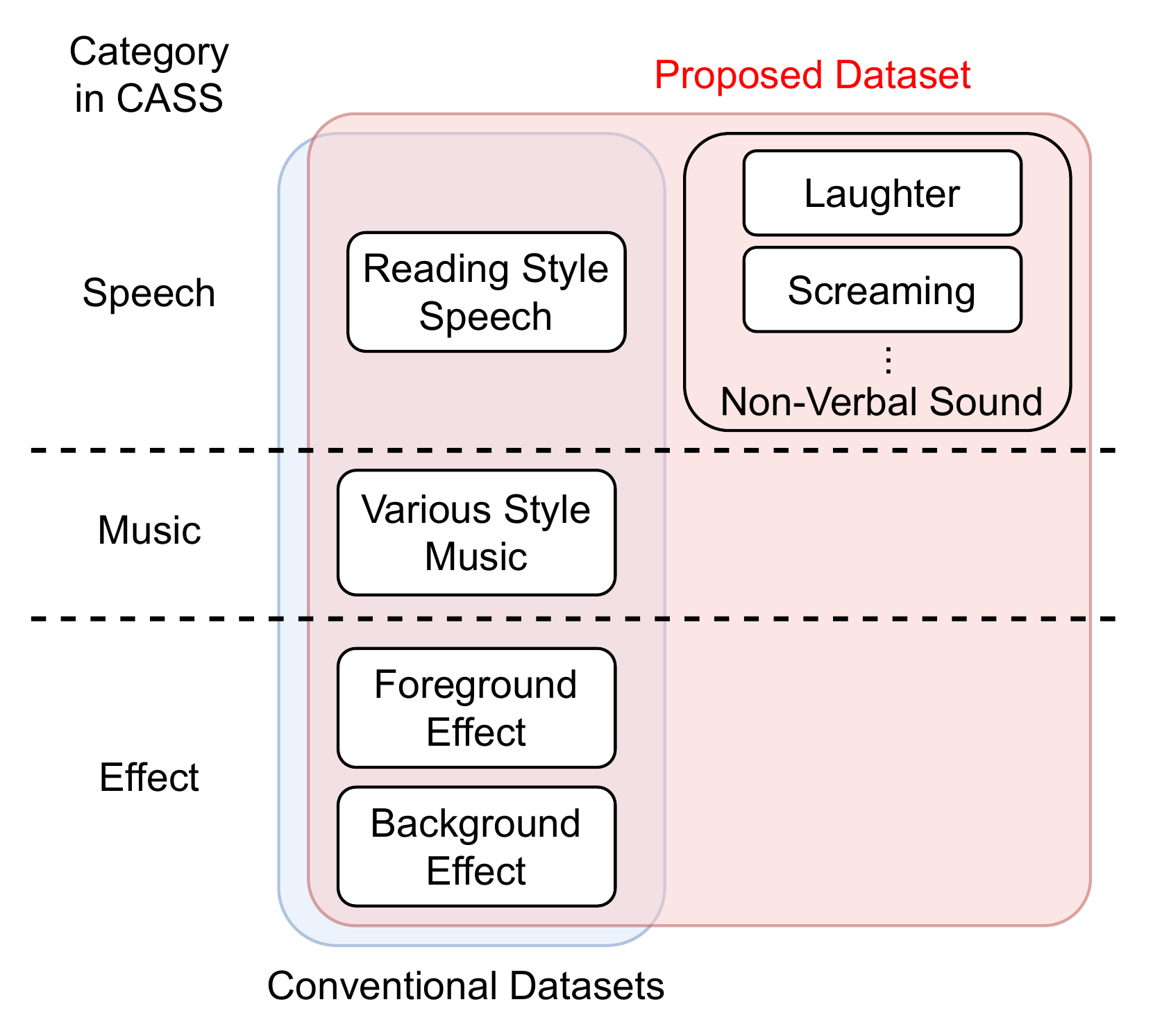}
    \caption{Comparision of conventional CASS datasets and proposed \textit{DnR-nonverbal}. Unlike conventional datasets such as DnR-v2, our dataset covers non-verbal sounds often observed in movie audio.}
    \label{fig:background/comparison}
\end{figure}

To address the limitations of existing CASS datasets, we propose a new dataset, \textit{DnR-nonverbal}, specifically designed to include non-verbal sounds as part of the speech stem.
The non-verbal clips are drawn from FSD50K and newly crawled from FreeSound to ensure a diverse range of non-verbal sounds.
We applied rule-based filtering and large language model (LLM)-based filtering to remove invalid clips.
Through experiments using synthetic movie audio, our findings illustrate that the current CASS model tends to extract non-verbal sounds as an effect stem due to the absence of non-verbal sounds in the conventional datasets.
Our dataset effectively addresses this issue by incorporating non-verbal sounds to bridge the gap between synthetic and actual movie audio.
Furthermore, the subjective evaluation using actual movie audio with non-verbal sounds shows that our dataset enables the CASS model to separate the vocal content more naturally and consistently.
Examples of clips and separation results are available at \url{https://tky823.github.io/hasumi2025dnr.github.io/}.

\section{CASS and conventional datasets}
\subsection{CASS formulation}
The mixing process of CASS is defined as follows:
\begin{align}
    \bm{y} = \bm{x}_{\mathrm{s}} + \bm{x}_{\mathrm{m}} + \bm{x}_{\mathrm{e}},
\end{align}
where $\bm{x}_{\mathrm{s}}$, $\bm{x}_{\mathrm{m}}$, and $\bm{x}_{\mathrm{e}}$ denote the monaural waveforms of speech, music, and effect respectively, and $\bm{y}$ is the mixture of stems.
$\bm{x}_{\mathrm{e}}$ can be defined as the mixture of foreground $\bm{x}_{\mathrm{f}}$ and background $\bm{x}_{\mathrm{b}}$ effects.
The CASS task is estimating $\bm{x}_{\mathrm{s}}$, $\bm{x}_{\mathrm{m}}$, and $\bm{x}_{\mathrm{e}}$, from observation $\bm{y}$.
As in \cite{petermann2022cocktail}, foreground and background effects are treated in a single stem as a separation target.

\subsection{Existing datasets}
\label{sec:datasets/dnr}
In the existing CASS datasets, each stem is designed by concatenating clips in corpora.

\noindent\textbf{DnR-v2} is a widely-known CASS dataset partially used for the CDX challenge~\cite{uhlich2024sound}.
In this dataset, music clips are sampled from FMA~\cite{defferrard2016fma}, which contains various genres of music.
The effects are drawn from FSD50K~\cite{fonseca2020fsd50k}, which contains various environmental sounds such as \textit{Vehicle}, \textit{Animal}, and \textit{Thunder}.
The source of speech is LibriSpeech~\cite{panayotov2015librispeech}, first used for automatic speech recognition (ASR) tasks.
Since LibriSpeech is built on an audiobook corpus, most speakers read aloud the text in a reading style.

\noindent\textbf{DnR-v3}~\cite{warcharasupat2024remastering} is another possible CASS dataset, an extension of DnR-v2.
Unlike DnR-v2, v3 contains speeches in languages other than English, which improves the diversity of the speech in terms of language families.
The sources of DnR-v3 are also a speech corpus and do not include non-verbal sounds.

\noindent\textbf{Speech-Music Datasets} also exist related to CASS.
The task targets only speech and music stems without effect.
Among them, LSX~\cite{petermann2023hyperbolic}, PodcastMix~\cite{schmidt2022podcastmix}, and JRSV~\cite{bai2024jointly} are the representative datasets.
These datasets use LibriSpeech, VCTK~\cite{veaux2017cstr}, or AISHELL-1~\cite{bu2017aishell} as speech stem.
Though various sources of speech corpus are used, all speeches are reading-style, similar to existing CASS datasets.

\section{DnR-nonverbal}
\subsection{Motivation}
In the actual movie audio, we can decompose $\bm{x}_{\mathrm{s}}$ as follows:
\begin{align}
    \bm{x}_{\mathrm{s}} = \bm{x}_{\mathrm{v}} + \bm{x}_{\mathrm{n}},
\end{align}
where $\bm{x}_{\mathrm{v}}$ and $\bm{x}_{\mathrm{n}}$ correspond to the waveforms of verbal and non-verbal sounds, respectively.
As described in Sec.~\ref{sec:datasets/dnr}, the conventional dataset contains only reading-style speech as verbal sounds (i.e., $\bm{x}_{\verbal} \approx \bm{x}_{\reading}$, where $\bm{x}_{\reading}$ is a reading-style speech) and omits non-verbal sounds (i.e., $\bm{x}_{\mathrm{n}}=\bm{0}$) from $\bm{x}_{\speech}$.
Though there is a discrepancy in $\bm{x}_{\speech}$ from the actual scenario, spontaneous speech can be extracted as a speech stem from the movie audio.
However, the assumption does not allow the model to extract non-verbal sounds as a speech stem.

For the CASS model to appropriately extract the non-verbal sound as speech, we propose \textit{DnR-nonverbal} dataset based on the DnR-v2 dataset.
As depicted in Fig.~\ref{fig:background/comparison}, our speech stem contains non-verbal sounds, such as laughter, screaming, and whispering voices, in addition to usual reading-style speeches, unlike the existing datasets.
In our dataset, each track is 60 seconds long.
Note that the difference between our dataset and DnR-v2 only lies in the speech stem.
We use the same mixing strategy for music and effect stems as DnR-v2.

Our definition of including non-verbal sounds as a speech stem is justified in representing vocal content within movie audio.
In movie audio, non-verbal sounds are uttered alongside linguistic speech and recorded on the same channel.
Splitting them into different stems or ignoring them is unnatural.
Rather, treating them as a single speech stem like our motivation is more reasonable for practical application.

\subsection{Collection of non-verbal sounds}
To include non-verbal sounds in the speech stem, we collect clips from FSD50K~\cite{fonseca2020fsd50k}.
Using the AudioSet~\cite{gemmeke2017audio} ontology, we gather clips with six voice-related tags: \textit{Laughter}, \textit{Whispering}, \textit{Crying\_and\_sobbing}, \textit{Screaming}, \textit{Sigh}, and \textit{Shout}, which are child tags of \textit{Human\_voice} in the ontology.
Note that, in DnR datasets, human-voice clips are removed from effect stems.

Though FSD50K is a valuable source for non-verbal sounds, it provides less than 400 clips except for \textit{Laughter}.
To increase the size of the dataset, we crawled additional clips from FreeSound\footnote{\url{https://freesound.org/}} via their API.
We collected clips that satisfy all of the following conditions:
\begin{itemize}
    \item
    The license is Creative Commons 0.
    \item
    The tags include at least one of \textit{screaming}, \textit{scream}, \textit{shout}, \textit{whispering}, \textit{whisper}, \textit{crying}, \textit{cry}, \textit{sobbing}, \textit{sob}, and \textit{sigh}.
    \item
    The clip is not created by mixing another clip of FreeSound and is not used to remix another one.
    \item
    The clip is not used for FSD50K since FSD50K was originally curated from FreeSound.
\end{itemize}

\vspace{-5pt}
\subsection{Filtering of collected clips}
\begin{figure}
    \centering
    \includegraphics[width=0.75\linewidth]{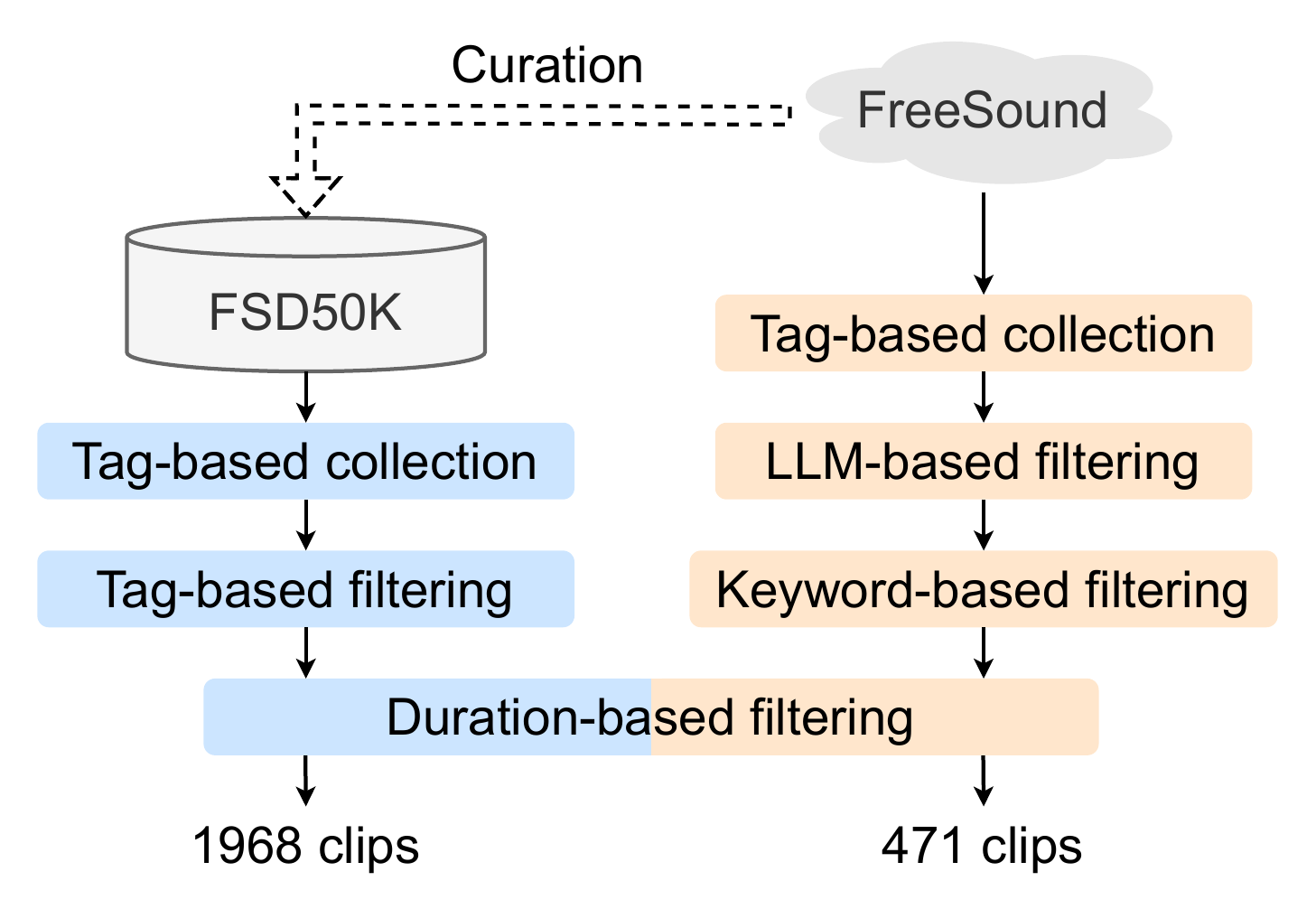}
    \caption{Filtering procedure to extract non-verbal clips from FSD50K dataset and FreeSound.}
    \label{fig:datasets/filtering-procedure}
\end{figure}
After collecting clips, we applied filtering to remove clips suspected of containing non-human voices or being too long.
Fig.~\ref{fig:datasets/filtering-procedure} shows the filtering procedure.

The tags on FSD50K clips are not exclusive and might have undesired tags as a speech stem.
We removed clips containing non-\textit{Human\_voice} descendant tags.
Even if the annotated tags are composed of only \textit{Human\_voice} descendant tags, we removed ones with \textit{Singing} tag to avoid the inclusion of music content.
These rule-based filterings yielded 1439, 135, and 395 clips for training, validation, and evaluation.

For the clips newly crawled from FreeSound, we utilized an LLM to enhance selection accuracy.
We made a prompt\footnote{The prompt is ``\textit{You have to decide whether the provided audio is available as a target of non-verbal speech extraction. You should determine the availability by guessing the given tags and description of the audio. The sample containing non-human sounds like applause, cars, and animals should be rejected. Singing voices should be rejected as well because they are treated as music. The low-quality or noisy sample should be rejected because it degrades the separation quality. Answer yes or no for availability as a non-verbal sound from tags \{tags\} and description `\{description\}'. Explanations are not allowed.}''} to roughly filter out clips with low quality or with non-human voice tags and input it to GPT-4o~\cite{hurst2024gpt}.
Only clips with a \textit{yes} response were retained. Despite LLM-based filtering, some clips still contained non-human sounds.
We removed them by keyword-based filtering and left 552 clips.

In the last step, we removed clips with more than 30 seconds to avoid one track being occupied by one clip.
After processing these filters, we obtained 1909, 135, and 395 clips as non-verbal sounds for training, validation, and evaluation, respectively.
Among them, 1968 clips are derived from FSD50K, and 471 clips are from FreeSound.
All FreeSound clips were included in the training set to avoid unexpectedly including invalid non-verbal sounds during evaluation.

\subsection{Mixing process in speech stem}
\label{sec:DnR-nonverbal/mixing-process}
\begin{figure}[t]
    \begin{algorithm}[H]
    \caption{Mix reading-style speech and non-verbal sounds}
    \label{alg:datasets/mixing-procedure}
        \begin{algorithmic}[1]
        \State \textbf{Definitions}
        \State \quad $L$: number of timesteps in track
        \State \quad $F$: sampling rate
        \State \quad $\ReadingSet$: list of reading-style speech clips
        \State \quad $\NonverbalSet$: list of non-verbal sound clips
        \State \quad $\ZTP$: zero-truncated Poisson distribution
        \State \quad $\tilde{\mathcal{G}}$: skew Gaussian distribution
        \State \quad $\lambda_{\reading}$, $\lambda_{\nonverbal}$: expected values of $\ZTP$
        \State \quad $\alpha$, $\sigma$: skew and scale parameters of $\tilde{\mathcal{G}}$
        \State \quad $A_{\speech}$: target loudness
        \State $M_{\reading} \sim \ZTP(\lambda_{\reading})$ \# number of reading-style speeches
        \State $M_{\nonverbal} \sim \ZTP(\lambda_{\nonverbal})$ \# number of non-verbal sounds
        \State $\mathcal{R}' \leftarrow \mathrm{sample}(\ReadingSet, M_{\reading})$
        \State $\mathcal{N}' \leftarrow \mathrm{sample}(\NonverbalSet, M_{\nonverbal})$
        \State $C\leftarrow\mathrm{shuffle}(\mathcal{R}' + \mathcal{N}')$ \# concatenate and shuffle
        \State $\bm{x}_{\speech} \leftarrow \bm{0} \in \mathbb{R}^{L}$
        \State $\tau\leftarrow 0$ \# current timestep
        \ForAll{$m = 1, \ldots, M_{\reading} + M_{\nonverbal}$}
            \State $\bm{c} \leftarrow \mathrm{pop}(C)$
            \If{$\tau + \mathrm{len}(\bm{c}) > L$}
                \State continue \# clip is too long
            \EndIf
            \State $d\sim\SkewNorm(\alpha,\sigma)$ \# sample silence duration
            \State $\ell \leftarrow \max(\lfloor Fd \rfloor, 0)$ \# duration to timesteps
            \State $\ell \leftarrow \min(\ell, L - \mathrm{len}(\bm{c}))$
            \State $\bm{x}_{\speech}[\tau:\tau+\ell] \leftarrow \bm{0}$
            \State $\tau\leftarrow\tau+\ell$
            \State $a \sim [A_{\speech} - 2, A_{\speech} + 2]$ \# sample loudness
            \State $\bm{c} \leftarrow \mathrm{rescale}(\bm{c}, a)$
            \State $\bm{x}_{\speech}[\tau: \tau + \mathrm{len}(\bm{c})] \leftarrow \bm{c}$
            \State $\tau\leftarrow\tau+\mathrm{len}(\bm{c})$
        \EndFor
        \If{$\bm{x}_{\mathrm{s}}$ does not contain a non-verbal sound clip}
        \State Retry from L11
        \EndIf
        \end{algorithmic}
    \end{algorithm}
\end{figure}
We follow the DnR-v2 dataset to create stems, except for the speech stem, for simulating movie audio.
Algorithm~\ref{alg:datasets/mixing-procedure} shows the mixing of reading style and non-verbal sounds in our dataset.

First, we sampled the numbers of reading-style speeches ($M_{\reading}$) and non-verbal sounds ($M_{\nonverbal}$) by zero-truncated Poisson distribution, setting expected values at $\lambda_{\reading} = 6$ and $\lambda_{\nonverbal} = 5$ , respectively.
The combined $M_{\reading} + M_{\nonverbal}$ clips are shuffled to make a list of clip candidates $C$.

Each clip is then popped from $C$ and preceded by a silence interval determined by a skew Gaussian distribution, with skew parameter $\alpha = 5$ and scale parameter $\sigma = 2$.
Clips that cannot fit into the track length are discarded to ensure all utterances are fully contained within the mixed track.

The volume based on loudness units full-scale (LUFS)~\cite{peaktowards} is randomly sampled by uniform distribution over $[A - 2, A + 2]$, where $A$ denotes the category-specific hyperparameter.
We set $A_{\speech} = -17$ for reading-style speech and non-verbal sounds, and $A_{\mathrm{m}}=-21$, $A_{\mathrm{f}}=-21$, and $A_{\mathrm{b}}=-29$ for music and effect stems.
These values are based on \cite{petermann2022cocktail}, which indicates the speech stem, including non-verbal sounds, is louder than other stems.

Finally, if the track does not contain non-verbal sounds, we drop it.
Following these steps, we prepared 1000, 50, and 100 tracks for training, validation, and evaluation.

\subsection{Dataset property}
\label{sec:datasets/property}
\begin{figure*}[t]
    \centering
    \begin{subfigure}[t]{0.3\textwidth}
        \includegraphics[width=\linewidth]{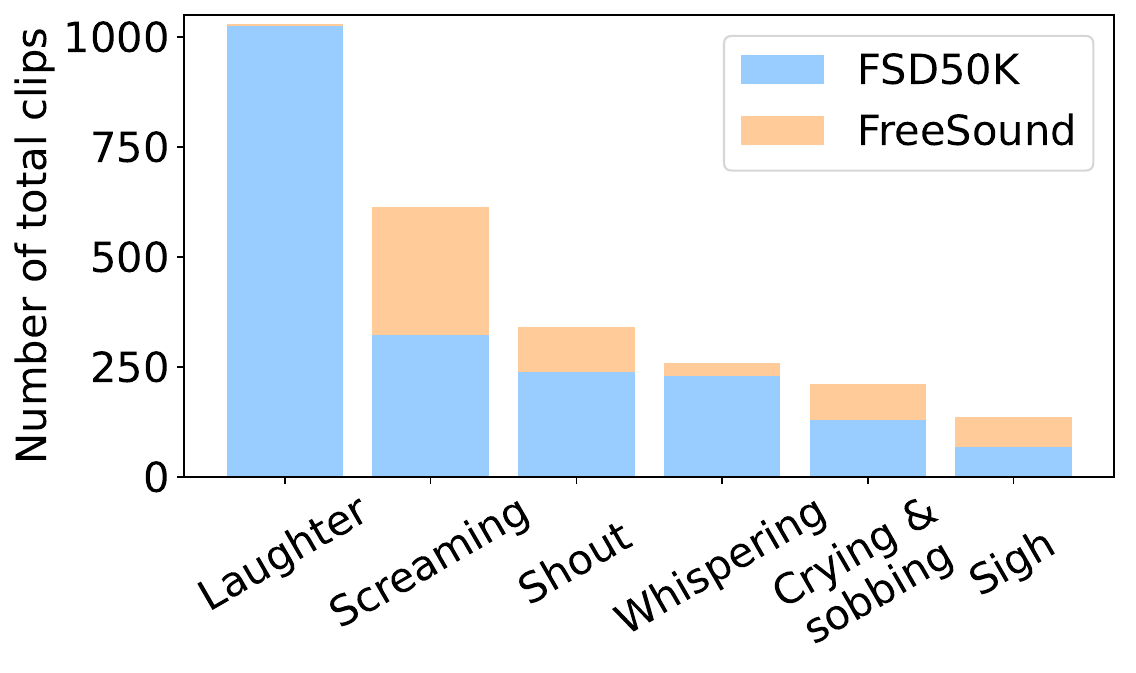}
        \caption{Count of clips per non-verbal sound tag in DnR-nonverbal.}
        \label{fig:datasets/tag-frequency}
    \end{subfigure}
    \hfill
    \begin{subfigure}[t]{0.3\textwidth}
        \includegraphics[width=\linewidth]{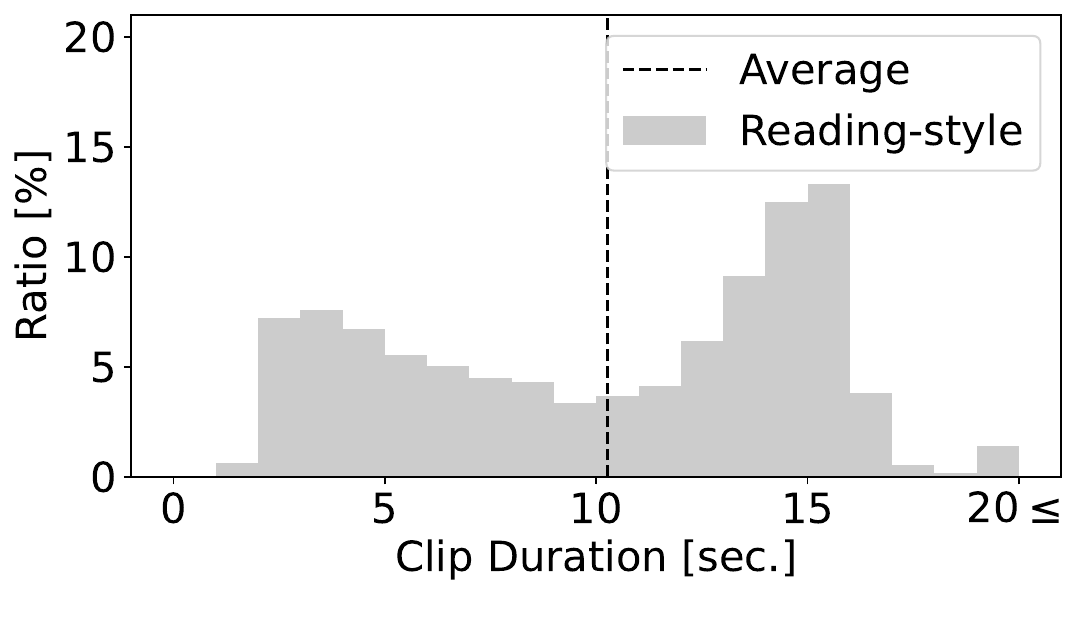}
        \caption{Distribution of clip durations in speech stems of DnR-v2.}
        \label{fig:datasets/nonverbal-duration-v2}
    \end{subfigure}
    \hfill
    \begin{subfigure}[t]{0.3\textwidth}
        \includegraphics[width=\linewidth]{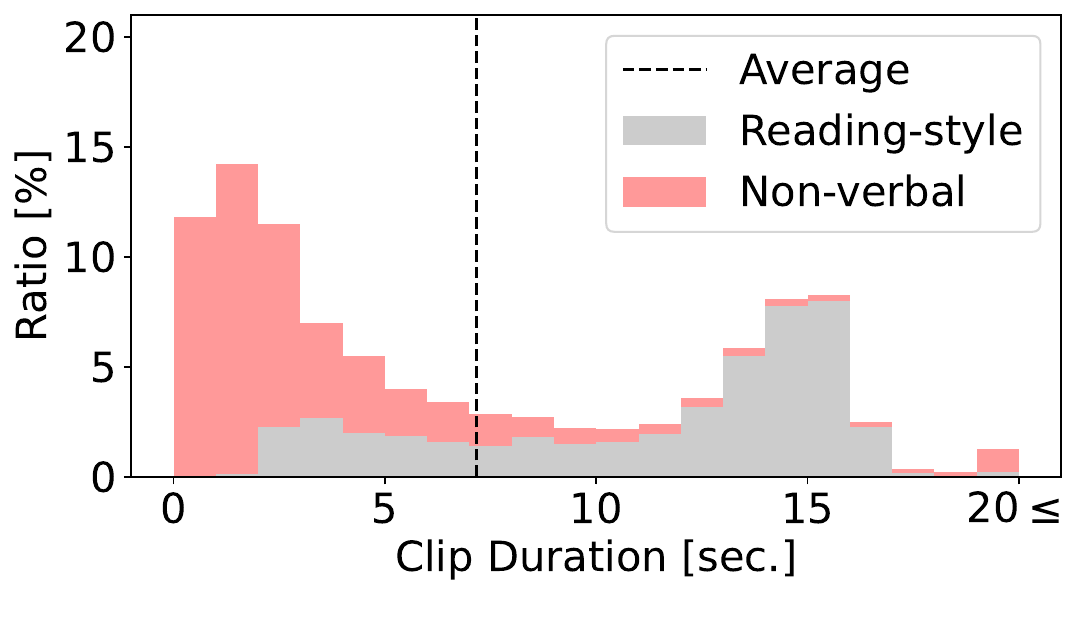}
        \caption{Distribution of clip durations in speech stems of DnR-nonverbal.}
        \label{fig:datasets/nonverbal-duration-proposed}
    \end{subfigure}
    \caption{Property of DnR-nonverbal dataset.}
    \label{fig:datasets/property}
\end{figure*}
Fig.~\ref{fig:datasets/property}(a) shows the number of clips per tag in DnR-nonverbal.
Each category contains at least 100 clips.
Among them, \textit{Laughter} contains about 1000 clips.
Furthermore, since most of \textit{Laughter} clips are composed of FSD50K, the sound is expected to be of high purity.
Other than \textit{Laughter}, the number of clips is less than 600, and a certain amount of clips derives from newly crawled FreeSound, which may contain the sounds from other categories.

Fig.~\ref{fig:datasets/property}(b) and \ref{fig:datasets/property}(c) show distributions of the durations of speech clips in DnR-v2 and DnR-nonverbal, respectively.
In both datasets, most clips are shorter than 15 seconds.
The non-verbal sounds in DnR-nonverbal are shorter than reading-style speech, lowering the average duration of speech clips.

\section{Experiments}
\subsection{Settings}
To evaluate the effectiveness of the proposed dataset, we conducted CASS experiments.
As a CASS model, we used BandIt~\cite{watcharasupat2023generalized} with long short-term memory~\cite{hochreiter1997long} backbone. 
The model is trained for 100 epochs by the sum of a frequency-domain mean-absolute-error (MAE)
loss and a time-domain MAE loss~\cite{luo2023music} using the Adam optimizer~\cite{kingma2014adam} with an initial learning rate of 0.001.
At every two epochs, we decayed the learning rate by multiplying 0.98.
We compared two training dataset conditions: DnR-v2 and DnR-v2 + DnR-nonverbal.
The batch size was set to 16 and randomly sampled 20k mini batches at every epoch following \cite{watcharasupat2023generalized}, regardless of the dataset size.
Each mixture is created by dynamic mixing~\cite{jeon2024does}.

After the training, the objective separation performance was measured by source-to-distortion ratio (SDR):
\begin{align}
    \mathrm{SDR} = 20 \log_{10}\frac{\lVert\bm{x}_{\mathrm{targ}}\rVert}{\lVert\bm{x}_{\mathrm{targ}} - \bm{x}_{\mathrm{est}}\rVert},
\end{align}
where $\bm{x}_\mathrm{targ}$ and $\bm{x}_\mathrm{est}$ denote target and estimated monaural waveforms.
We used the model that marked the best performance on the validation set for evaluation.

\vspace{-2pt}
\subsection{Non-verbal sound extraction performance}
\label{sec:experiments/performance}
\begin{table}[t]
    \caption{SDR scores of speech ($\bm{x}_{\mathrm{s}}$) and effect ($\bm{x}_{\mathrm{e}}$) stems in the evaluation set of DnR-nonverbal. $\bm{x}_{\mathrm{r}}$ and $\bm{x}_{\mathrm{n}}$ are reading-style speech and non-verbal sounds included in $\bm{x}_{\mathrm{s}}$, respectively. $\hat{\bm{x}}_{\cdot}$ denotes a stem estimated by the model trained on DnR-v2. The conventional dataset makes the model misallocate non-verbal sounds as the effect stem.}
    \label{tab:experiments/degradation}
    \centering
    \begin{tabular}{cc|c}
        \hline
        $\bm{x}_{\mathrm{est}}$ & $\bm{x}_{\mathrm{targ}}$ & SDR [dB] \\ \hline
        $\hat{\bm{x}}_{\mathrm{s}}$ & $\bm{x}_{\mathrm{r}} + \bm{x}_{\mathrm{n}} (=:\bm{x}_{\speech})$ & $5.62$ \\
        $\hat{\bm{x}}_{\mathrm{e}}$ & $\bm{x}_{\mathrm{e}}$ & $2.54$ \\ \hline
        $\hat{\bm{x}}_{\mathrm{s}}$ & $\bm{x}_{\mathrm{r}}$ & $6.52$ \\
        $\hat{\bm{x}}_{\mathrm{e}}$ & $\bm{x}_{\mathrm{e}} + \bm{x}_{\mathrm{n}}$ & $7.08$ \\ \hline
    \end{tabular}
\end{table}
\begin{table}[t]
    \vspace{-2pt}
    \caption{SDR scores for evaluation set of DnR-nonverbal.}
    \label{tab:experiments/performance}
    \centering
    \begin{tabular}{c|ccc|c}
        \hline
        Training Dataset(s) & Speech & Music & Effect & All \\ \hline
        DnR-v2 & $5.62$ & $4.33$ & $2.54$ & $4.16$ \\
        DnR-v2 + & \multirow{2}{*}{$\bm{9.30}$} & \multirow{2}{*}{$\bm{4.79}$} & \multirow{2}{*}{$\bm{5.23}$} & \multirow{2}{*}{$\bm{6.44}$} \\
        DnR-nonverbal & & & & \\ \hline
    \end{tabular}
\end{table}
To reveal that the model trained by the conventional dataset finds extracting non-verbal sounds as a speech stem challenging, we evaluated the performance by changing the definition of the non-verbal sound category.
Table~\ref{tab:experiments/degradation} shows the SDR scores when the non-verbal sound is defined as speech and when it is defined as an effect stem.

The SDR scores are improved by only changing the definition of the non-verbal sounds from speech to effect.
This result indicates that the model trained by the conventional dataset extracts the non-verbal sounds as the effect rather than the speech, even though there are no non-verbal sounds in training datasets.
The model may treat neither reading-speech nor music content as effects.

\subsection{Overall performance on DnR-nonverbal}
Table~\ref{tab:experiments/performance} shows the SDR scores in the evaluation set of DnR-nonverbal.
From the results, the model trained by DnR-v2 + DnR-nonverbal shows significantly higher scores in speech and effect stems.
This indicates that the CASS model can recognize the non-verbal sound as a speech stem by mixing non-verbal sounds into the reading-style speech.
In addition, the score of the music stem is slightly improved as a side effect.

\subsection{Subjective evaluation by actual movie audio}
\begin{table}[t]
    \caption{Result of A/B tests on speech extraction performance using actual movies.}
    \label{tab:experiments/subjective-evaluation}
    \centering
    \begin{tabular}{ccc}
        \hline
        DnR-v2 wins & on par & DnR-v2 + DnR-nonverbal wins \\ \hline
        $4.2\%$ & $18.8\%$ & $\bm{76.9\%}$ \\ \hline
    \end{tabular}
\end{table}
Though our evaluation set contains non-verbal sounds in the speech stem, a discrepancy remains from the actual movie audio.
To investigate the separation quality of realistic movie audio, we conducted A/B tests using 20 tracks from \url{Movieclips.com}, each 6 seconds long and containing non-verbal sounds.
13 raters were asked to watch a video with the original sound and two sound-edited versions.
One version uses an extracted speech stem from the model trained on DnR-v2, while the other is by the model trained on DnR-v2 + DnR-nonverbal.
Then, they were asked which sound was more natural and consistent as the extraction result of the voice of the actors.

Table~\ref{tab:experiments/subjective-evaluation} shows the results of the A/B tests.
The model trained by DnR-v2 + DnR-nonverbal scores significantly higher due to its ability to extract non-verbal sounds.
This observation indicates that there indeed exists a mismatch between the conventional datasets and the actual movie audio.
Our dataset demonstrates its effectiveness in actual movie audio, suggesting a heightened potential for the trained CASS model to be a practical audio processing tool in the filmmaking and editing industry.

During the subjective evaluation, we found a small negative effect with the proposed dataset: the dataset could cause the model to mistake the voice of an animal for screaming.
This problem will be alleviated by introducing a vision model that considers the context of the movie.

\section{Conclusion}
In this paper, we highlighted the underlying issue of the conventional CASS dataset: non-verbal sounds are excluded in any stems, which led the trained CASS model to treat expressive voice as an effect stem.
To address this issue, we built a new dataset containing non-verbal sounds named \textit{DnR-nonverbal}.
Our dataset contains non-verbal sounds such as laughter, screaming, and whispering as a speech stem.
From the objective evaluation, adding our dataset to the conventional datasets improves the performance of the CASS model in synthetic movie audio.
Furthermore, we showed that our dataset is also effective in the actual movie audio containing various non-verbal sounds.
We hope our dataset will help with tasks such as query-based audio source separation and audio captioning, as well as CASS.

\end{document}